\documentclass[]{spie}  
\usepackage[]{graphicx}
\title {Atmospheric transparency in the optical and near IR range above the Shatdzhatmaz summit} 
\author{Olga Voziakova\supit{a}
\skiplinehalf
\supit{a}Sternberg Astronomical Institute, Universitetsky prosp., 13, Moscow, Russia
}
\authorinfo{E-mail: ovoz@sai.msu.ru}
\begin{document} 
\maketitle 
\begin{abstract}
The study of atmospheric extinction based on the MASS data
has been carried out using the classical photometric pairs method.
The extinction in V band can be estimated at $0.^m19$.
The water vapour content has been derived from GPS measurements.
The median value of PWV for clear nights is equal to 7.7 mm.
\end{abstract}
\section{Atmospheric extinction in the MASS spectral band}

\label{sec:mass}  
MASS-DIMM instrument has been working on Shatdzhatmaz summit for over
three years.
Being developed for scintillation measurements, MASS is a proper 
photoelectric photometer. 
We use star fluxes measured by MASS in the greater
apertures C and D for the atmospheric extinction
investigation\cite {MASSDIMM_MNRAS2007}.
We do not apply the Bouguer method because it implies the stability of atmospheric
extinction during the night. Unfortunately, such conditions generally do not
take place, particularly in the Caucasus. 
To allow for extinction variations we employ a method based on photometric pairs.
Unlike the classical one, we use instrumental magnitudes of extinction stars. 
It was necessary, therefore, to make a list of stars with
determined instrumental magnitudes $m_{mass}$in the MASS photometrical
band ($\lambda_{eff}$ = 481 nm). 

Seventeen stars were selected by following criteria:
non-variable, approximately A0 spectrum, high culminating in the Caucasus.
We observed these stars from August 2008 to October 2010.
Each clear night they were measured every 1.5 hours on equal
and different air masses.
To derive the $m_{mass}$ only best nights with small transparency variations
have been selected. 
The system of 1065 equations with 171 unknowns  has been worked out. 
The unknowns are $m_{mass}$ of 17 standard stars and the effective extinctions of 154 nights.
As the zero point of our photometric system we have chosen a star $\gamma$ Gem
of A0IV type with $B-V=0$. 
We assign to $m_{mass}$ of this star its $m_B$ value from BS catalogue. 
The system has been solved by the least squares method. Table ~\ref{tab:stars}
lists the derived magnitudes. Their errors average out to $0.^m003$, which  introduce
an error of about $0.^m01$ into the extinction calculation.

\begin{figure}
\begin{center}
\begin{tabular}{cc}
\includegraphics [angle = -90]{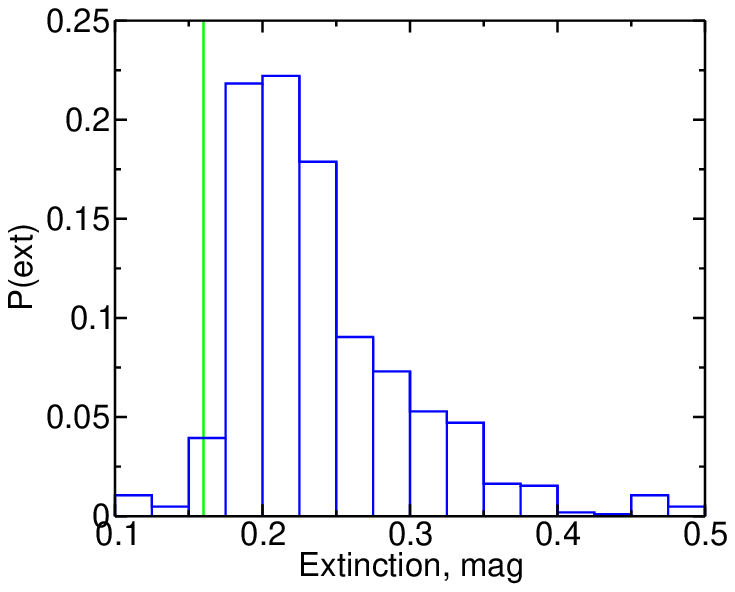} &
\includegraphics [angle = -90]{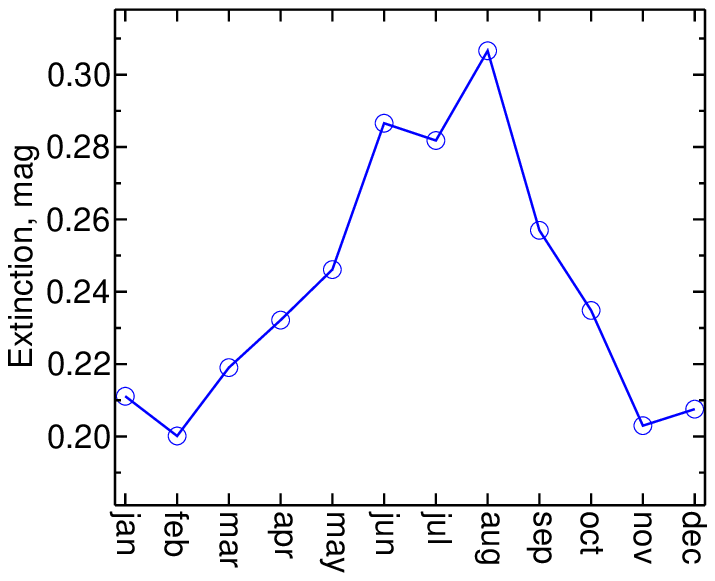} \\
\end {tabular}
\end {center}
\caption[ext] 
{\label{fig:ext}
{\it Left:} Distribution of atmospheric extinction in the $MASS$ band.
Vertical green line represents theoretical Rayleigh extinction for
Shatdzhatmaz. {\it Right:}
Seasonal behaviour of median atmospheric extiction. }
\end {figure}

Using magnitudes from Table ~\ref{tab:stars} we obtain the fine extinction variation
during the night. 
In Fig.~\ref{fig:ext} (left) the distribution of 1040 extinction
measurements is presented . The median is $0.^m23$ (1st quartile, $0.^m20$; 3rd quartile, $0.^m27$).
The theoretical Rayleigh extinction for Shatdzhatmaz summit (2100 meters
above sea level) is $0.^m16$ in the $MASS$ band and $0.^m13$ in $V$ band\footnote
{The atmospheric transparency was modelled using MODTRAN\cite {1987MODTRAN} for A0V type.}.
The total extinction in $V$ band can be estimated at $0.^m19$ on the assumption that aerosol is
neutral.

In Fig.~\ref{fig:ext} (right) the seasonal behaviour of the median extinction is shown.
One can see that minimum extinction is observed in winter. In the
same period there are many clear nights.
\begin{table}
\caption{The list of photometric stars} 
\label{tab:stars}
\begin{center}       
\begin{tabular}{|l|l|l|c|c|l|} 
\hline
 star    & $\alpha_{2000}$ & $\delta_{2000}$ & $m_{mass}$ & $\sigma_m$ & type \\
\hline
$\beta   $ Ari & 01 54 38 & +20 48 29 &    2.721 &   0.003  &        A5V\\
$\beta   $ Tri & 02 09 33 & +34 59 14 &    3.084 &   0.003  &        A5II\\
$\eta    $ Tau & 03 47 29 & +24 06 18 &    2.841 &   0.003  &        B7IIIe\\
$\beta   $ Tau & 05 26 18 & +28 36 27 &    1.605 &   0.002  &        B7III\\
$\gamma  $ Gem & 06 37 43 & +16 23 57 &    1.930 &   0.000  &        A0IV\\
$\beta   $ UMa & 11 01 51 & +56 22 57 &    2.342 &   0.002  &        A1V\\
$\delta  $ Leo & 11 14 07 & +20 31 25 &    2.612 &   0.003  &        A4V\\
$\beta   $ Leo & 11 49 04 & +14 34 19 &    2.181 &   0.005  &        A3V\\
$\gamma  $ UMa & 11 53 50 & +53 41 41 &    2.423 &   0.003  &        A0Ve\\
$\epsilon$ UMa & 12 54 02 & +55 57 35 &    1.749 &   0.002  &        A0pCr\\
$\eta    $ UMa & 13 47 32 & +49 18 48 &    1.780 &   0.003  &        B3V\\
$\zeta   $ Dra & 17 08 47 & +65 42 53 &    3.116 &   0.003  &        B6III\\
$\delta  $ Her & 17 15 02 & +24 50 21 &    3.158 &   0.003  &        A3IV\\
$\gamma  $ Lyr & 18 58 57 & +32 41 22 &    3.221 &   0.003  &        B9III\\
$\zeta   $ Aql & 19 05 25 & +13 51 48 &    2.989 &   0.004  &        A0Vn\\
$\alpha  $ Cep & 21 18 35 & +62 35 08 &    2.562 &   0.003  &        A7V\\
$\alpha  $ Peg & 23 04 46 & +15 12 19 &    2.466 &   0.005  &        B9Vs\\
\hline
\end{tabular}
\end{center}
\end{table} 

\section{Atmospheric transparency in the near IR range} 
\label{sec:IR}
As well known, the transparency of the atmosphere in the IR range is
determined by the water vapour content. We estimate the amount of precipitable
water using zenith wet delay of GPS signal\cite {Bevis94}.
Since the Caucasus is a rather wet region, the accuracy of this method is
sufficient.
The GPS data were obtained from a GPS receiver\footnote {The GPS receiver was installed by Sternberg
Astronomical Institute in the frame of the geodynamics research.}, located on the territory of the Solar Station at a distance of 800 meters from the
MASS-DIMM tower. The meteo station is 100 meters
away from GPS. The meteo data were kindly provided
by the Kislovodsk station of the
A.M.Obukhov Institute of Atmospheric Physics (Russian Academy of Sciences).

\begin{figure}
\begin{center}
\begin{tabular}{cc}
\includegraphics [angle = -90]{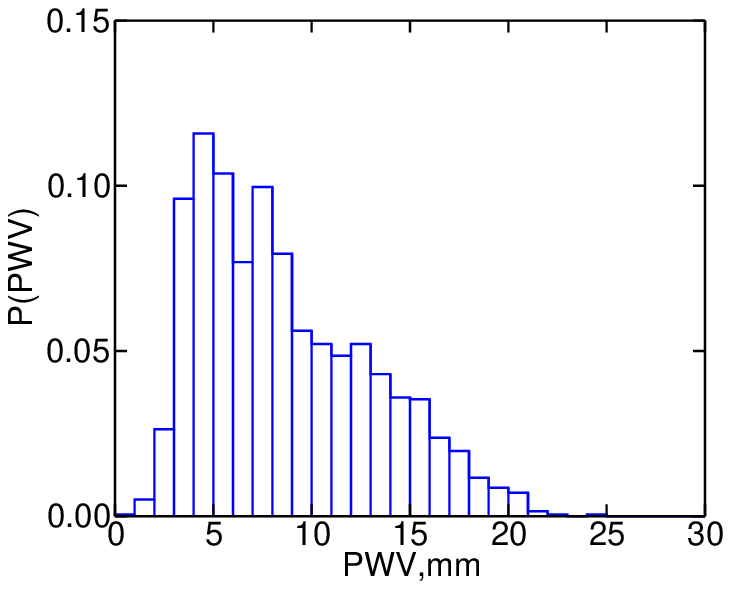} &
\includegraphics [angle = -90]{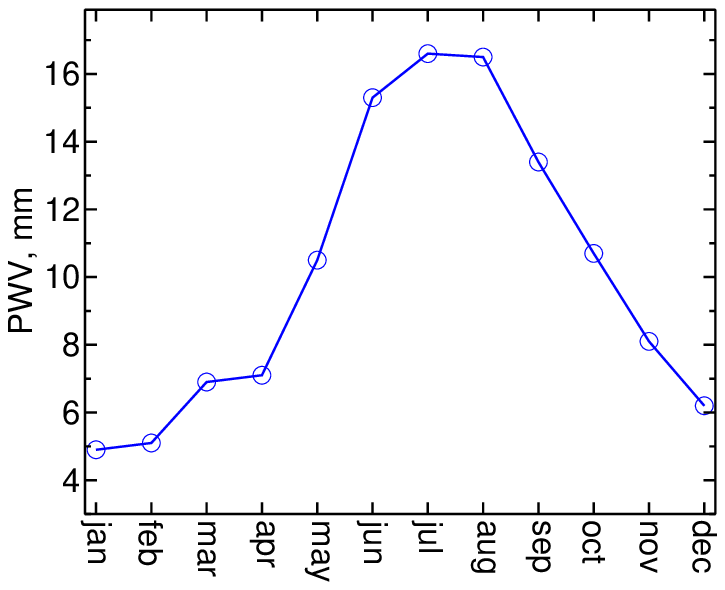}\\

\end{tabular}
\end{center}
\caption[pw] 
{ \label{fig:pw}
{\it Left:} Distribution of precipitable water vapour on clear nights.
{\it Right:} The seasonal behaviour of precipitable water vapour.}
\end {figure}

The median value of PWV estimated on the total GPS data set is equal to 9.1
mm (1st quartile, 6.1 mm; 3rd quartile, 13.9 mm). Distribution of PWV on clear nights
is given in Fig.~\ref{fig:pw} (left).
The median of this sample is 7.7 mm (1st quartile, 5.1 mm; 3rd quartile, 11.8 mm). The 7.7 mm corresponds to atmospheric
extinction for A0V-star in J band of $0.^m26$, in H and K bands of
$0.^m12$.
The water vapour content displays an obvious seasonal cycle
(Fig.~\ref{fig:pw}, right), most
likely due to seasonal course of temperature.
It is seen that winter and spring are the best seasons for infrared
observations.

\section{
Some attempts to predict atmospheric conditions from ground-based
meteorological observations. } 
\label{sec:predict}
It would be useful to find a way to predict the night sky conditions by means of
simple weather observations. To clear sky monitoring we use the Boltwood cloud sensor located at the
Solar Station. The meteo data are provided by the Kislovodsk station of the A.M.Obukhov Instutute of
Atmospheric Physics and our automatic site monitor (ASM)\cite {ASM_MNRAS2010}.

\begin{figure}
\begin{center}
\begin{tabular}{cc}
\includegraphics [angle = -90]{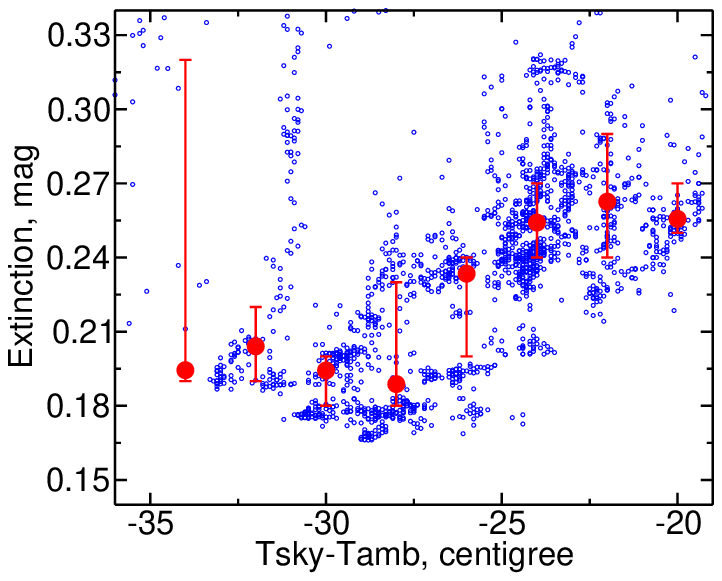} &
\includegraphics [angle = -90]{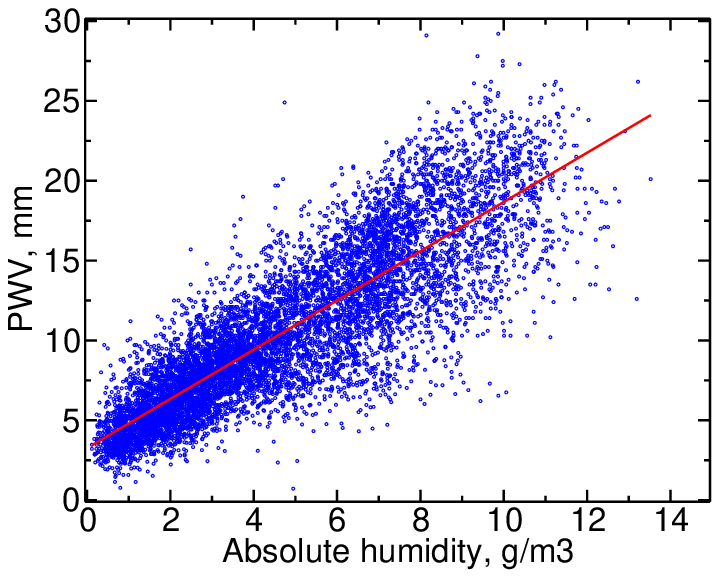} \\
\end {tabular}
\end {center}
\caption[meteo]
{ \label{fig:meteo}
{\it Left:} The cloud sensors's sky temperature versus atmospheric extinction
in the $MASS$ band. Red line is the median extinction in 2-centigree
bins.
{\it Right:} Absolute humidity versus PWV. The linear fit is shown as a red
line (see text).
}
\end {figure}

The relationship between sky temperature and atmospheric extinction
in the $MASS$ band is shown in Fig.~\ref{fig:meteo} (left).

We have calculated surface absolute humidity using the dew point temperature and air temperature.
Absolute humidity is the density of water vapour $D_w (kg/m^3)$.
From the ideal gas law equation:
    $$D_w = \frac {P_w} {(T_c+273.15) R_w}\mbox{,}$$
where $P_w$ is water vapour pressure in $Pa$, \\
$R_w=461.5$ $(J \, kg^{-1} \, K^{-1})$ is the specific gas constant for water
vapour, and\\
$T_c$  is the temperature in degrees Celsius.\\
From Magnus-Tetens approximation\cite{barenbrug1974psychrometry}:
$$P_w = 611 \cdot 10^{ \frac{7.5 T_{dc}} {237.7+T_{dc}} }\mbox{,}$$
where $T_{dc}$  is the dew point temperature in degrees Celsius.

Fig.~\ref{fig:meteo} (right) represents surface absolute humidity versus column of precipitable water vapour.
This relationship can be fitted by the linear regression:
$PWV = 3.2514 + 1.5419 \cdot D_w$,
where $PWV$ is measured in $mm$, $D_w$ is the surface absolute humidity in
$g/m^3$. Rms error is 2.62 mm.

Figurs~\ref{fig:meteo} show a quite wide scatter, which means that we have
to look for additional parameters of these relations.

\section {Conclusions}
\label{sec:concl}
MASS data can be useful not only for atmospheric profiling but also for
transparency measuring, e.g., for
realtime observation scheduling by photometric stability criterion.

In spite of rather high humidity the observations in the near IR
range may be carried out in the Caucasus region.

\acknowledgments     
I am grateful to Dr. V. G. Kornilov for valuable advices, to
Obukhov Institute of Atmospheric Physics and I. A. Senik for kindly  provided
meteo data, to A. O. Ivlev (Research and Production Corporation
"Precision Instrumentation Systems") for MODTRAN interface and to gravimetric team of
Sternberg Astronomical Institute for assistance with GPS data processing.
\bibliography{report}   
\bibliographystyle{spiebib}   

\end{document}